\documentclass[aps,twocolumn,superscriptaddress,preprintnumbers]{revtex4}
\usepackage{epsfig}
\usepackage{graphicx,amssymb,epsfig,amsmath}
\usepackage{color}
\usepackage{float}
\restylefloat{table}

\newcommand{\be}{\begin{eqnarray}}
\newcommand{\ee}{\end{eqnarray}}

\newcommand{\tr}[1]{\text{tr}\left[#1\right]}
\newcommand\bigforall{\mbox{\large $\mathsurround=0pt\forall$}}

\begin{document}

\title{Quantum advantage for distributed computing without communication}

\author{\L{}. Czekaj}

\affiliation{Faculty of Mathematics, Physics and Informatics, Gda\'nsk University, 80-952 Gda\'nsk,Poland}
\affiliation{National Quantum Information Centre in Gda\'nsk, 81-824 Sopot, Poland}

\author{M. Paw\l{}owski}

\affiliation{Faculty of Mathematics, Physics and Informatics, Gda\'nsk University, 80-952 Gda\'nsk,Poland}
\affiliation{National Quantum Information Centre in Gda\'nsk, 81-824 Sopot, Poland}

\author{T. V\'{e}rtesi}
\affiliation{Institute for Nuclear Research, Hungarian Academy of Sciences, H-4001 Debrecen, P.O. Box 51,
Hungary}

\author{A. Grudka}

\affiliation{Faculty of Physics, Adam Mickiewicz University,
  Umultowska 85, 61-614 Pozna\'{n}, Poland}

\author{M. Horodecki}

\affiliation{Faculty of Mathematics, Physics and Informatics, Gda\'nsk University, 80-952 Gda\'nsk,Poland}
\affiliation{National Quantum Information Centre in Gda\'nsk, 81-824 Sopot, Poland}
\author{R. Horodecki}

\affiliation{Faculty of Applied Physics and Mathematics, Gda{\'n}sk
  University of Technology, 80-952 Gda{\'n}sk, Poland}
\affiliation{National Quantum Information Centre in Gda\'nsk, 81-824 Sopot, Poland}

\begin{abstract}
Understanding the role that quantum entanglement plays as a resource in various information processing tasks is one of the crucial goals of quantum information theory.
Here we propose a new perspective for studying quantum entanglement: distributed computation of functions without communication between nodes.
To formalize this approach, we propose {\it identity games}.
Surprisingly, despite of no-signaling, we obtain that non-local quantum strategies beat classical ones
in terms of winning probability for identity games originating from certain bipartite and multipartite functions.
Moreover we show that, for majority of functions, access to general non-signaling resources boosts success probability two times in comparison to classical ones, for number of outputs large enough.

\end{abstract}

\maketitle


Since famous Shor algorithm \cite{shor_prime_factor}, a basic branch of quantum information science is devoted to search for quantum advantage in computing. In particular a lot of effort was devoted to distributed computation -- mostly in terms of communication complexity.
A strictly related domain is a huge "industry" of Bell inequalities, which are actually  an
instance of distributed computation.

\begin{figure}
    \centering
        \includegraphics[scale=0.35]{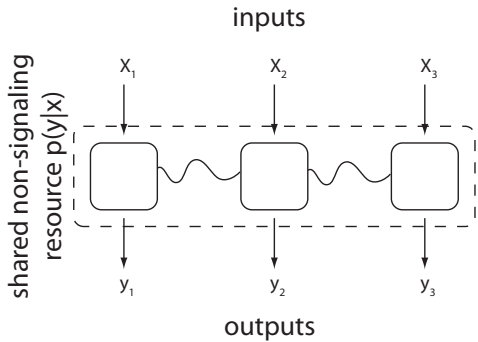}
    \caption[Distributed computing without communication]
    {{\it Distributed computing without communication (Id games)} Players share non-signaling resource, here understood as a correlations $p(\vec{y}|\vec{x})$.
    Each player receives private message $x_k\in\{ 0,1,\ldots,m_i-1 \}$ from referee. Messages are distributed according to the uniform probability distribution. Based on the input $x_k$ and his part of shared resource, player $k$ computes the output $y_k\in\{ 0,1,\ldots,m_o-1 \}$.
No communication is allowed between the players during the game.
Players have to compute some total function $f$, i.e. they win the game if their outputs fulfill $\vec{y}=f(\vec{x})$.}
\label{fig:setup}
\end{figure}

So far, the paradigm of {\it distributed computation} assumed communication between the nodes.
On the one hand the communication was directly quantified and the cost of communication was measured as in communication complexity problems (see \cite{buhrman_review} for review). On the other hand some global inputs/outputs processing was performed by a referee \cite{linden_distributed}.
Here we focus on computation without communication.

Namely, we want to address the following question: {\it is quantum
mechanics superior to classical theory regarding distributed
computation of total function, if no communication between nodes
is allowed?} We answer this question introducing and studying {\it
Identity games} (in short {\it Id games}, the origin of the name
will be clarified later) which may be viewed as special type of
Bell inequalities~\cite{CHSHBellIntro}. The setup is depicted in
FIG.~\ref{fig:setup} and described with more details in further
part of the paper.

We should here stress the difference between the problem we pose and two existing results/approaches.
The first one is  ``guess your neighbour input'' (GYNI) game \cite{gyni}. GYNI game may be viewed as a distributed computation without communication with global constraints imposed on the input set. In contrast, in our approach,
we consider all inputs. Secondly, in  \cite{Francuzi} the communication complexity problem for two parties was translated
into scenario, where the goal of one of the parties is to compute some function without communication, but conditioned on an event that other party will not abort. This conditioning is a form of communication, while in our case there is no communication whatsoever.

Intuitively, computing of some non-trivial distributed function requires signaling.
Hence there should not be a difference not only between quantum and classical theories,
but also non-signaling theory should not give advantage.
As an example we may refer to GYNI game where all inputs are allowed. In that case there is no advantage for non-signaling theories.




In this paper we show that quantum systems are superior to
classical ones for bipartite and multipartite scenarios. We find
that, for the simplest bipartite systems (i.e. each player has
binary input and output) non-signaling theory does not offer
advantage. However, if we move to more complicated bipartite
setups, we obtain that quantum strategies beat
classical ones.




Before we present our results, we discuss how the approach
proposed in this paper relates to other frameworks for studying
non-locality, in particular to Bell inequalities.

The value of Bell expression is the average probability that some
function $f(\vec{x})$ of parties' inputs $\vec{x}$ is equal to
some function $g(\vec{y})$ of their outputs $\vec{y}$ \footnote{In
some instances $g$ may depend also on $\vec{x}$ or the average
weighted~\cite{i3322}.}. The most well studied type of Bell
inequalities are so-called XOR games \cite{xor}. There, the
function $g$ is simply XOR of the outcomes of all parties.
Celebrated CHSH \cite{CHSH} inequality is the most well known
example of this type.
The main advantage of XOR games is that they allow to change a more complex function $g$ into computing XOR's.
As we said, the goal of distributed computation is to reach a point
when no further processing is required.
Because in this case $g$ is just the identity it is why we call the inequalities of this type {\it Id games} (i.e. in the case of total functions).

Strangely this case has been left almost untouched with only one, as mentioned above, exception which is GYNI game \cite{gyni}.
As contrary to GYNI game where partial function is computed, here we focus on the total functions $f$. 
To our knowledge there are no examples of Bell inequalities of type of Id games for total functions.

In analogy to communication complexity problems
\cite{buhrman_review}, these Id games may by viewed as functional
games where only one output value $\vec{y}$ is a valid answer for
given input $\vec{x}$. This is in contrast to the relational games
(e.g. XOR game) where more than one answer is correct for given
input $\vec{x}$.

As in communication complexity problems, Id games are intended to
capture an advantage from using non-local resources in distributed
computation. In contrast with communication complexity, we do not
measure advantage in terms of communication cost. Actually, no
communication between parties is allowed in our approach. We focus
on the probability that players return correct output.





This paper is structured as follows: Firstly we formally define Id games and resources that can be used to play them.
Then we show various examples with quantum advantage for bipartite scenario and discuss the gap between classical and non-signaling resources. Next we move to tripartite scenario. At the end we show that for a vast majority of the functions the classical resources are practically useless while the no-signalling ones allow for strictly better results. We conclude the paper with a discussion of our results and directions for future research.

In the Method section we present numerical examples, some
statistics describing number of nontrivial Id games and other
details.


\section*{RESULTS}

\subsection*{Identity games}

We study Bell inequalities in terms of nonlocal games. We consider
$n$-player games. Each player $k$ receives as an input a private
message from referee  $x_k\in\{0,1,\ldots,(m_i-1)\}$ and returns
to him an output $y_k\in\{0,1,\ldots,(m_o-1)\}$. $m_i$ and $m_o$
denote numbers of possible inputs and outputs respectively.
Inputs are distributed according to some joint probability distribution $q(\vec{x})$ on vector $\vec{x}\in\{0,1,\ldots,(m_i-1)\}^n$
consisting
of messages sent to each player.
The goal for players is to achieve maximal value of functional called winning probability:
\begin{equation}
\omega = \sum_{\vec{x},\vec{y}}\delta\big(f(\vec{x})=g(\vec{y})\big)q(\vec{x})p(\vec{y}|\vec{x}),\label{eq:winprob_general}
\end{equation}
where $f(\vec{x})$ and $g(\vec{y})$ are some functions of inputs and outputs  and define the game,  $p(\vec{y}|\vec{x})$ is conditional probability distribution describing the strategy.

In the preparation step, players are informed on the input
distribution $q(\vec{x})$ and functions $f$ and $g$. Then they
establish common strategy $p(\vec{y}|\vec{x})$ based on the shared
resources (classical, quantum, non-signaling). During the game, no
communication is allowed between the players. Hence they make
their decision on the base of their own inputs and shared
resources. We will denote maximal winning probability for
classical, quantum and non-signaling resources as $\omega^*_{cl},
\omega^*_{q}, \omega^*_{ns}$ respectively. The game is non-trivial
if $\omega^*_{cl} < \omega^*_{ns}$.

In this paper we look for non-trivial Id games which are defined
by the requirement that $g$ is an identity function. Furthermore
we consider only the case where $g(\vec{x})=1/m_i^n$. That leads
to winning probability in the form:
\begin{equation}
\omega = \sum_{\vec{x},\vec{y}}\delta(f(\vec{x})=\vec{y}) \frac{1}{m_i^n} p(\vec{y}|\vec{x}).\label{eq:winprob_id}
\end{equation}

\subsection*{Strategies}

We describe strategies used by players in terms of generalized
probabilistic theories. Players share $n$-partite resources. Each
player perform some measurement $x_k$ on his part of the resource.
The measurements are performed simultaneously {and their results
are distributed according to joint conditional probability
$p(\vec{y}|\vec{x})$. The} following theories, characterized by
the conditions imposed on the $p(\vec{y}|\vec{x})$, are important
for our purposes: (i) classical, (ii) quantum and (iii)
non-signaling. In classical theory $p(\vec{y}|\vec{x})$ is a
mixture of classical local probabilities according to unknown
hidden parameter $\lambda$. Two party resource
$p(\vec{y}|\vec{x})$ is of the form
\begin{equation}
p_{cl}(y_1,y_2|x_1,x_2) = \sum_\lambda p(\lambda)
p( y_1 | x_1 , \lambda )
p( y_2 | x_2 , \lambda ).
\end{equation}
For quantum theory, conditional probability distribution has to be
reproduced by local measurements (described by projectors
$P^{(1)}_{x,y},P^{(2)}_{x,y}$) performed on the shared quantum
state $\rho_{12}$:
\begin{equation}
p_{q}(y_1,y_2|x_1,x_2) = \tr{ \rho_{12} P^{(1)}_{x_1,y_1}\otimes P^{(2)}_{x_2,y_2}}.
\end{equation}
In non-signaling theories, the only condition is that $p(\vec{y}|\vec{x})$ cannot signal between players. In formal way this condition
is expressed for player $1$ as
\begin{equation}
\mathop{\bigforall}_{x_2,x_2'}
\sum_{y_2} p_{ns}(y_1,y_2|x_1,x_2) =
\sum_{y_2} p_{ns}(y_1,y_2|x_1,x_2')
\end{equation}
and similarly for player $2$.

To identify functions which lead to non-trivial Id games, we
grouped the functions into equivalence classes invariant under local operations or players reordering. Namely we treat functions $f_1$
and $f_2$ as equivalent if $f_2$ might be obtained from $f_1$ by composition of the following operations:
(i) input relabelling; (ii) output relabelling;
(iii) output conditioning on local input: player $k$ returns as an output $h_k(y_k,x_k)$;
(iv) players reordering.
These operations enabled for significant reduction of problem complexity.
Then for one representative function from each equivalence class we obtained
$\omega^*_{cl}$ and $\omega^*_{ns}$ by proper optimisation of~\eqref{eq:winprob_id}:
(i) $p_{cl}(\vec{y}|\vec{x})$ is convex combination of extreme strategies which are easy to enumerate, we calculate $\omega^*_{cl}$ by
maximisation of~\eqref{eq:winprob_id} over that set; (ii) $\omega^*_{ns}$ was obtained from linear programming with constraints
imposed
by no signaling conditions.

For non-trivial Id games we analysed performance of quantum
strategies. Since we were not able to provide analytical results,
we focused on upper bounding $\omega^*_{q}$ numerically. For this
purpose we used semidefinite programming (SDP) an approach
introduced in \cite{SDP1,SDP2}. Beside the upper bound, for some
cases we provided also concrete examples of quantum strategy
offering advantage over classical ones. For all
considered bipartite scenarios these advantage is equal to upper
bound obtained from SDP.


In the rest of the paper, for simplicity, we take binary inputs
($m_o=2$). The results obtained in this and following section can
be generalized to any $m_o>2$ in a straightforward way.



\subsection*{Two-player games}

In the scenarios with binary input (i.e. $m^{(k)}_i=2;m_o=2$),
there are $4^4=256$ functions, however none of them leads to
non-trivial Id game.

\textit{Three inputs per player} - The simplest setup where we can
find non-trivial Id games and quantum advantage is the case where
each player obtains one of the three input symbols
($m^{(k)}_i=3;m_o=2$). There are $4^9=262144$ different functions
in that setup which reduce to $2162$ equivalence classes.
There are $256$ equivalence classes for which $\omega^*_{cl} < \omega^*_{ns}$. Together they contain $196992$ functions.
For all of these classes $\omega^*_{cl} = 4/9\approx0.4444$ and $\omega^*_{ns}=1/2$.

We analyzed some of these classes in more detail looking for
quantum strategies which may have advantage over classical ones.
Interestingly we found that non-signaling strategies winning the
games are equivalent to using PR-Boxes \cite{PRBoxIntro} and
an explicit example of optimal quantum strategy
corresponds to the CHSH Bell experiment \cite{CHSH}.

We calculated upper bound for $\omega_{q}$ using 2-nd
level hierarchies of SDP \cite{SDP1}. The highest bound is about
$5.2\%$ better than classical strategy and reads
$0.4675$. In comparison non-signaling strategy has
$12.5\%$ advantage. On the other hand we found that for some
classes the bound obtained from SDP is equal (with respect to the
numerical precision) to $\omega^*_{cl}$.

The example of function with the highest SDP bound is:
\begin{table}[H]
\centering
\begin{tabular}{r|ccc}
$x_2\backslash x_1 (y_2,y_1)$&0&1&2\\
\hline
0&0,0&0,0&0,0\\
1&0,0&1,1&1,1\\
2&0,1&0,1&1,1\\
\end{tabular}
\end{table}
Each cell refers to the output of function for given input. First
and second player's inputs are in the columns and rows
respectively.




Optimal classical strategy for this game is obtained when both players return $0$ all the time.
Optimal non-signaling strategy is analogical to using PR-Box:
\begin{equation}
p(y_1,y_2|x_1,x_2) =
    \begin{cases}
   1/2 & \text{if } y_1\oplus y_2 = ( x_2 = 2 \wedge x_1 = 0 )\\
   1/2 & \text{if } y_1\oplus y_2 = ( x_2 = 0 \wedge x_1 = 2 )\\
   0   & \text{otherwise}
  \end{cases}
\end{equation}

Quantum strategy which is optimal for this game
(i.e. its attains the SDP bound \cite{SDP1}) looks as follows:
Player 1 outputs deterministically $0$ when he gets $x_1=0$ as an
input. Otherwise he relabels his input $\{1\}\rightarrow
\{1\},\{2\}\rightarrow \{0\}$ and plays XOR game with optimal
quantum strategy. Player 2 performs analogically with relabeling
$\{1\}\rightarrow \{0\},\{2\}\rightarrow \{1\}$. Note that, when
one of the players return $0$ deterministically, another one
returns $0$ or $1$ completely random. Success probability
$\omega_q$ for this strategy is $(1+\frac{3}{2}+
\frac{\sqrt{2}+2}{2})/9\approx 0.4675$.



However, if we consider another Id game (see Methods section for
details), we find that the quantum maximum $\omega^*_{q}$ is
attained by using non-maximally entangled two-qubit states.

On the other hand, Id games turn out to be useful as dimension
witnesses as well, i.e., they are able to witness Hilbert space
dimension \cite{dimwit}. In an Id game given in Methods section
the players must share at least three dimensional component spaces
for maximum quantum violation $\omega^*_{q}$ to happen.

We also found that the SDP bound is equal to classical limit for
functions which were symmetric according to players exchange. As
an example we present the following symmetric function:
\begin{table}[H]
\centering
\begin{tabular}{r|ccc}
$x_2\backslash x_1 (y_2,y_1)$&0&1&2\\
\hline
0&0,0&0,0&1,0\\
1&0,0&1,1&1,1\\
2&0,1&1,1&0,0\\
\end{tabular}
\end{table}
An optimal non-signaling strategy is analogical to using PR-box:
\begin{equation}
p(y_1,y_2|x_1,x_2) =
    \begin{cases}
   1/2 & \text{if } y_1\oplus y_2 = ( x_2 = 2 \wedge x_1 \neq 2 ) \\
   0   & \text{otherwise}
  \end{cases}
\end{equation}

The above feature that quantum strategies do not offer advantage
over classical ones turns out to be true if the players receives
four inputs each (instead of three). However, we found
counterexample for the case of five inputs per player (see
Methods).

\textit{Four inputs per player} - Let us now move to the scenario
where each of the two players may receive 4 inputs ($m^{(k)}_i=4
;m_o=2$). This scenario is computationally too costly to classify
all $4^{16}$ different functions and make an exhaustive search for
quantum violations. However, it is worthy to highlight a few
particular Id games for which we find $\omega^*_{cl} < \omega^*_q
< \omega^*_{ns}$:

(i) \textit{Addition game.} Let us define the Id game with the
following function
\begin{equation}
2y_2 + y_1 = x_1 + x_2 \text{ mod } 4,
\end{equation}
where $x_1, x_2$ are assumed to take values in $\{0,1,2,3\}$
whereas $y_1,y_2\in\{0,1\}$. The $\vec y=(y_2,y_1)$ above encodes
in two bits the result of adding two base-4 integers $x_1$ and
$x_2$ in a modulo 4 arithmetic. For this game, we have the success
probabilities, $\omega^*_{cl} = 3/8 = 0.375$, $\omega^*_q = (2 +
\sqrt 2)/8 \approx 0.4268$, and $ \omega^*_{ns} = 1/2$. Hence,
$\omega^*_q$ and $\omega^*_{ns}$ beat the classical limit
$\omega^*_{cl}$ by about $13.81\%$ and $33.33\%$, respectively.
Both values represent considerable improvement over the 3-input
case discussed previously. The quantum maximum $\omega^*_q \approx
0.4268 $ is attained by using a maximally entangled two-qubit
state and co-planar measurements which are presented in the
Methods section.

(ii) \textit{A facet-defining game.} It turns out that there
exists a 4-input Id game, which defines a facet of the Bell local
polytope in the scenario of four binary inputs per party. This
game is equivalent to the Bell inequality $I_{4422}^6$ defined by
the paper of Brunner and Gisin~\cite{BG08}. See Methods section
for an explicit construction of this game.




\subsection*{Three-player games}

Here we consider three-player games with binary inputs
($m^{(k)}_i=m_o=2$). There are $8^8=16777216$ different functions
which reduce to $5876$ equivalence classes. We found $68$
equivalence classes (with $34176$ functions together) for which
$\omega^*_{cl} < \omega^*_{ns}$. According to classification of
extremal non-signaling strategies \cite{3pnsclass}, for most of
equivalence classes, $\omega^*_{ns}$ is achieved by decomposable
strategies (i.e. strategies which may be decomposed into PR-box on
2 parties and local deterministic box on the remaining party).
More details on the classification of winning strategies are
presented in the Methods section.

Here we focus on one game where we can also provide an example of quantum strategy with advantage over classical one. The function $f$ for this game may be written as:
\begin{eqnarray}
y_1 & = &(\bar{x}_1\wedge\bar{x}_2)\oplus\bar{x}_3\label{eq:f3y0}\\
y_2 & = &\bar{x}_3\\
y_3 & = &0\label{eq:f3y2}.
\end{eqnarray}

In the optimal classical strategy all players return $0$ all the time which leads to
$\omega^*_{cl}=0.375$. Optimal non-signaling strategy decomposes into PR-box
shared between player 1 and 2 and deterministic strategy used by player $3$ (he always returns $0$).
For that strategy we have $\omega^*_{ns}=0.5$.

Inspired by decomposability of optimal non-signaling strategy, we
propose the following quantum strategy: player $1$ and player $2$
apply quantum strategy with maximal success probability for XOR
game, i.e. the one optimizing $y_2\oplus y_1 = \bar{x}_2\wedge
\bar{x}_1$. Player $3$ uses deterministic strategy: he returns $0$
all the time. Success probability achieved by quantum strategy for
2-player XOR game is $\cos^2\frac{\pi}{8}$. It is easy to see from
\eqref{eq:f3y0}-\eqref{eq:f3y2} that $y_2\oplus y_1 =
\bar{x}_2\wedge \bar{x}_1$. Only one of two possible outputs
winning XOR games is valid for the discussed function, and hence
for this quantum strategy we get $\omega_q=\frac12
\cos^2\frac{\pi}{8} = 0.42677 > \omega^*_{cl}$. This value may be
compared with the bound obtained from SDP which reads $0.42683$
($1+AB$ hierarchy of \cite{SDP1,SDP2}).


\subsection*{Generic advantage of Id games with multiple outcomes}

We now argue, that for any number of players, for large enough
number of outputs, the no-signaling theories beat the classical
ones generically. Specifically, let us define as $M_{cl}(\omega)$
($M_{ns}(\omega)$) to be the number of functions for $n$ parties,
with $m_i=m_o=m$, for which the probability of successful
implementation within classical (non-signaling) theory is
$\omega$.  We show that {\it for any number of parties $n$ the
ratio $\frac{M_{cl}(2^{1-n})}{M_{ns}(2^{1-n})}$ goes to zero for
increasing $m$.} The proof is deferred to the Methods section.

\section*{METHODS}

\subsection*{Two-player games}

\textit{Id game using partially entangled states} - We show an Id
game which allows higher winning probability using partial
entangled states than maximally entangled states of any dimension.
The game is as follows:
\begin{table}[H]
\centering
\begin{tabular}{r|cccc}
$x_2\backslash x_1 (y_2,y_1)$&0&1&2\\
\hline
0&0,1&1,1&1,0\\
1&0,0&0,1&1,1\\
2&0,1&1,0&0,1\\
\end{tabular}
\end{table}
\noindent with $\omega^*_{cl} = 4/9$ and $\omega^*_{nl} = 1/2$.

Using maximally entangled states (of any dimension),
$\omega^*_{q+} = 4.0178/9$. This value has been certified by using
the SDP method introduced in Section~4.2 of \cite{zoo}. However,
the maximum using general quantum resources is $\omega^*_q =
4.1224/9$, which saturates the SDP upper bound of \cite{SDP1}. In
fact, $\omega^*_q$ can be attained by using a partially entangled
two-qubit state. Hence, the winning probability attainable with
maximally entangled states (of any dimension) is strictly smaller
than the one using non-maximally entangled qubits.

\textit{Id game as a dimension witness} - We present here an Id
game which allows higher quantum violation if more than
two-dimensional systems are considered. Hence, this Id game also
gives an example to a dimension witness~\cite{dimwit}: Maximum
quantum violation does not happen in two dimensional systems. The
players have to conduct measurements on at least three dimensional
systems for maximum quantum violation to happen. The game is as
follows:
\begin{table}[H]
\centering
\begin{tabular}{r|cccc}
$x_2\backslash x_1 (y_2,y_1)$&0&1&2\\
\hline
0& 0,1 & 1,1 & 1,0\\
1& 0,1 & 1,1 & 1,1\\
2& 0,1 & 1,0 & 1,0\\
\end{tabular}
\end{table}
\noindent with $\omega^*_{cl} = 4/9$ and $\omega^*_{nl} = 1/2$.
Actually, no quantum violation can be observed for two dimensional
systems, however, $\omega^*_{q} = 4.1547005/9$ by using three
dimensional systems. This value is certified by SDP hierarchy
\cite{SDP1} as well.

\textit{Id game performing addition} - The function defining the
game can be written as
\begin{equation}
2y_2 + y_1 = x_1 + x_2 \text{ mod } 4,
\end{equation}
where $x_1,x_2$ take values in $\{0,1,2,3\}$ and
$y_1,y_2\in\{0,1\}$, which is represented by the table
\begin{table}[H]
\centering
\begin{tabular}{r|cccc}
$x_2\backslash x_1 (y_2,y_1)$&0&1&2&3\\
\hline
0&0,0&0,1&1,0&1,1\\
1&0,1&1,0&1,1&0,0\\
2&1,0&1,1&0,0&0,1\\
3&1,1&0,0&0,1&1,0\\
\end{tabular}
\end{table}
This can be further written as the following Bell functional:
\begin{equation}
\label{Ii}
I_{\text{add}}=\frac{-I_2^{(0,1;0,1)}+I_2^{(2,3;0,1)}+I_2^{(0,1;2,3)}-I_2^{(2,3;2,3)}+16}{64},
\end{equation}
with the CHSH game \cite{CHSH}: $I_2^{(i,j;m,n)}=-\langle A_i
B_m\rangle + \langle A_i B_n\rangle + \langle A_j B_m\rangle +
\langle A_j B_n\rangle$, where $\langle A_i B_m\rangle =
p(a=b|i,m)-p(a\neq b|i,m)$, where $a,b$ take values in $\{0,1\}$.
Due to Tsirelson \cite{Tsi80}, the quantum maximum for $I_2$ is
$2\sqrt 2$, from which an upper bound of $(4(2\sqrt 2)+16)/64 =
(2+\sqrt 2)/2\approx 0.4268$ for $I_{\text{add}}$ easily follows.
Interestingly, this upper bound can be obtained by an explicit
quantum strategy. Just take the qubit observables $A_0=\sigma_x$,
$A_1=\sigma_z$, $B_0=(\sigma_x-\sigma_z)/2$, and
$B_1=(-\sigma_x-\sigma_z)/2$ along with the Bell state
$|\phi^+\rangle = (|00\rangle + |11\rangle)/\sqrt 2$ (where
$\sigma_x, \sigma_z$ refer to Pauli matrices). This provides
$2\sqrt 2$ for the quantity $-I_2^{(0,1;0,1)}$ in Eq.~(\ref{Ii}).
Then choose the rest of the observables as $A_2 = -A_0$,
$A_3=-A_1$, $B_2=-B_0$, and $B_3=-B_1$. These choices maximize the
other three CHSH quantities as well saturating the upper bound
$\omega^*_q = (2+\sqrt 2)/2\approx 0.4268$ for $I_{\text{add}}$.

\textit{A facet-defining Id game} - The function to be considered
is as follows:
\begin{table}[H]
\centering
\begin{tabular}{r|cccc}
$x_2\backslash x_1 (y_2,y_1)$&0&1&2&3\\
\hline
0&0,1&1,0&0,0&1,0\\
1&0,1&1,1&0,1&1,1\\
2&0,0&1,1&0,0&1,0\\
3&0,0&1,0&1,0&0,0\\
\end{tabular}
\end{table}
This table translates to the following Bell functional:
\begin{equation}
\label{Iii} I_{\text{facet}}=\frac{I_{4422}^6+16}{64},
\end{equation}
where
\begin{align}
I_{4422}^6 =& -I_2^{(1,0;1,0)}+I_2^{(1,0;3,2)}+I_2^{(2,3;1,0)}+I_2^{(2,3;3,2)}\nonumber\\
           &-2\langle A_3B_0\rangle-2\langle A_3B_2\rangle+2\langle A_2\rangle+2\langle A_3\rangle
\end{align}
is equivalent up to input/output relabellings with the
$I^6_{4422}$ expression listed in the appendix of
Ref.~\cite{BG08}.

The maximum quantum violation is attained by using the Bell state
$|\phi^+\rangle$ and observables $A_i=\vec u_i\cdot\vec\sigma$ and
$B_i=\vec v_i\cdot\vec\sigma$, $i=0,1,2,3$, where
$\vec\sigma=(\sigma_x,\sigma_y,\sigma_z)$ is the vector of Pauli
matrices, and Alice and Bob's respective Bloch vectors $\vec u_i$,
$\vec v_i$  are given as follows:
\begin{align}
\vec u_0 =& \left(\sqrt{1-p^2}, 0, -p\right)\nonumber\\
\vec u_1 =& \left(-\sqrt{1-p^2-\frac{12}{15}},\sqrt{\frac{12}{15}},-p\right)\nonumber\\
\vec u_2 =& \left(2\sqrt{1-p^2-\frac{12}{60}},\sqrt{\frac{12}{60}},2p\right)\nonumber\\
\vec u_3 =& -\vec u_2,
\end{align}
and
\begin{align}
\vec v_0 =& \left(0, 0, 1\right)\nonumber\\
\vec v_1 =& \left(-\frac{\sqrt{5}}{3},0,-\frac{2}{3}\right)\nonumber\\
\vec v_2 =& \left(\sqrt{\frac{8}{9}-q^2},-q,\frac{1}{3}\right)\nonumber\\
\vec v_3 =& \left(\sqrt{\frac{5}{9}-q^2},q,-\frac{2}{3}\right),
\end{align}
with $p=0.408248$ and $q=0.730297$. With these settings, we get
$\omega^*_q = 0.403093$ for $I_{\text{facet}}$, which agrees with
the upper bound on level $1+AB$ of the SDP hierarchy. Note that
the Bloch vectors (both for Alice and Bob) span the full
three-dimensional space. Actually, the measurements attaining
$\omega^*_q$ cannot be brought to a co-planar form and
consequently require the use of complex numbers.

\textit{Symmetric Id game with quantum advantage} - We present
here a symmetric Id game for five inputs per player which offers
quantum advantage against classical strategies.
\begin{table}[H]
\centering
\begin{tabular}{r|ccccc}
$x_2\backslash x_1 (y_2,y_1)$&0&1&2&3&4\\
\hline
0& 1,1 & 1,0 & 0,0 & 0,0 & 1,1\\
1& 0,1 & 0,0 & 0,1 & 1,1 & 1,1\\
2& 0,0 & 1,0 & 0,0 & 1,1 & 0,1\\
3& 0,0 & 1,1 & 1,1 & 0,0 & 0,0\\
4& 1,1 & 1,1 & 1,0 & 0,0 & 0,0\\
\end{tabular}
\end{table}
We found that the quantum maximum is $\omega^*_q = 10.2950849/25$
(certified by SDP \cite{SDP1} as well) by performing measurements
on a 2-qubit singlet state. The classical bound, on the other
hand, is $\omega^*_{cl}=10/25$.

\subsection*{Three-player games}

Here we provide more detailed description of the numerical results
obtained for 3-player Id games. We provide some statistics for
$\omega^*_{cl}$ and $\omega^*_{ns}$ and classification of optimal
non-signaling strategies according to~\cite{3pnsclass} for these
classes where $\omega^*_{ns}>\omega^*_{cl}$.

Number of equivalence classes for which given $\omega^*_{cl}$ is
obtained:
\begin{table}[H]
\begin{tabular}{l|c}
$\omega^*_{cl}$&$\#$\\
\hline
0.25&   45\\
0.375&  23
\end{tabular}
\end{table}
Number of equivalence classes for which given $\omega^*_{ns}$ is
obtained:
\begin{table}[H]
\begin{tabular}{l|c}
$\omega^*_{ns}$&$\#$\\
\hline
0.275 & 1\\
0.28125 &   1\\
0.291667 &  11\\
0.3 & 1\\
0.3125 &    30\\
0.333333 &  1\\
0.4375 &    21\\
0.5 &   2
\end{tabular}
\end{table}
The difference between $\omega^*_{ns}$ and $\omega^*_{cl}$ with
the number of equivalence classes where it occurs:
\begin{table}[H]
\begin{tabular}{l|c}
$\Delta_a=\omega^*_{ns}-\omega^*_{cl}$&$\#$\\
\hline
0.025   &1\\
0.03125 &1\\
0.041667    &11\\
0.05    &1\\
0.0625  &51\\
0.083333    &1\\
0.125   &2
\end{tabular}
\end{table}
Similar list for relative differences between $\omega^*_{ns}$ and
$\omega^*_{cl}$:
\begin{table}[H]
\begin{tabular}{l|c}
$\Delta_r=\omega^*_{ns}/\omega^*_{cl} - 1$&$\#$\\
\hline
0.1 &1\\
0.125   &1\\
0.166666667 &21\\
0.166668    &11\\
0.2 &1\\
0.25    &30\\
0.333332    &1\\
0.333333333 &2
\end{tabular}
\end{table}
We do not present statistics for SDP bound of $\omega^*_{q}$ since
numerical inaccuracy make it hard to group them into classes.

Since non-signaling strategies which achieve $\omega^*_{ns}$ for
given $f$ are extremal points of non-signaling polytope, we may
apply classification from~\cite{3pnsclass} to them. We found that
optimal strategies belong to the following classes:
\begin{table}[H]
\begin{tabular}{l|c}
class&$\#$\\
\hline
2&  53\\
19& 1\\
25& 6\\
29& 6\\
31& 1\\
33& 1
\end{tabular}
\end{table}
There may be also strategies from the other classes since we just
checked to which class belongs the strategy obtained from linear
programming. Strategies from class $2$ may be decomposed into
PR-box on $2$ parties and local deterministic box on the remaining
party. Strategies from classes $25$ and $29$ are optimal for GYNI
game.

Decomposable strategies were discussed in the main part of the
paper. Here we give an example of optimal strategy which belongs
to class $25$.

The function of the game is given by equations:
\begin{eqnarray}
y_1 & = & \bar{x}_3\\
y_2 & = & \bar{x}_3\\
y_3 & = & (\bar{x}_3\wedge\bar{x}_1) \vee (x_3\wedge\bar{x}_2)
\end{eqnarray}
Classical and non-signaling strategies optimal for that game
achieve $\omega^*_{cl}=1/4$, $\omega^*_{ns}=1/3$ respectively
which give the gap $\Delta_a = 1/12$, $\Delta_r=1/3$. SDP bound
for $\omega^*_{q}$ is $0.260746$

In optimal classical strategy once again all players return $0$
all the time while optimal non-signaling strategy has the form (we
bolded the entries which win the game):
\begin{table}[H]
\begin{tabular}{l|cccccccc}
$x_3x_2x_1\backslash y_3y_2y_1$&000&001&010&011&100&101&110&111\\
\hline
000&1/3&0&1/3&0&0  &0&0&{\bf 1/3}\\
001&1/3&0&0&{\bf 1/3}&0  &0&1/3&0\\
010&1/3&0&1/3&0&0  &0&0&{\bf 1/3}\\
011&1/3&0&0&{\bf 1/3}&0  &0&1/3&0\\
100&0&0&1/3&1/3&{\bf 1/3}  &0&0&0\\
101&0&0&1/3&1/3&{\bf 1/3}  &0&0&0\\
110&{\bf 1/3}&0&0&1/3&0  &0&1/3&0\\
111&{\bf 1/3}&0&0&1/3&0  &0&1/3&0
\end{tabular}
\end{table}

\subsection*{Generic advantage of Id games for no-signalling theories}

We will now argue, that for any number of players, for large
enough number of outputs, the no-signaling theories beat the
classical ones generically. Specifically, let us define as
$M_{cl}(\omega)$ ($M_{ns}(\omega)$) to be the number of functions
for $n$ parties, with $m_i=m_o=m$, for which probability of
successful implementation within classical (non-signaling) theory
is $\omega$.  We will show:

{\it For any number of parties $n$ the ratio
$\frac{M_{cl}(2^{1-n})}{M_{ns}(2^{1-n})}$ goes to zero for
increasing $m$. }

To prove the claim, let us first note that, for any function $f$,
non-signalling theories allow $\omega_{ns}$ to be at least
$2^{1-n}$ (so that $M_{ns}$ is actually the number of all
functions). This is achieved by a box defined as follows. Let
$(f_1,...,f_n)=f(x_1,...,x_n)$. Define the probability
distribution of the box to be \be
P(y_1,...,y_n|x_1,...,x_n)=\left\{
\begin{array}{cc}
2^{1-n} & \bigoplus_{k=1}^n y_k = \bigoplus_{k=1}^n f_k \\
0 & \bigoplus_{k=1}^n y_k \neq \bigoplus_{k=1}^n f_k
\end{array}
\right. \ee It is straightforward to check that this box is
non-signalling and the probability that $\vec{y}=f(x_1,...,x_n)$
is exactly $2^{1-n}$ for any input. =

Let us now turn to classical theory. Obviously for any function we
can have $\omega_{cl}\geq 2^{-n}$. It is realized by a strategy in
which all parties return random outputs regardless of their
inputs. There are $M=2^{nm^n}$ possible functions $f$. This number
comes from the fact that for each function we need to specify the
outcome for each of $n$ players for any combination of $m^n$
possible inputs. We want to show that for a vast majority of
functions $f$ the classical resources (shared randomness and local
computation) do not allow to reach a significantly higher
probability. More precisely, our aim is to find (an upper bound
on) $M_c(\omega_{cl})$ -- the number of functions with the average
success probability of the best classical strategy larger or equal
to particular $\omega_{cl}$.


One nice property of the classical average success probability is
that there exists optimal strategy which is deterministic. (cf.
\cite{Fine}) -- namely each party applies function $y_k$ to his
input $x_k$, $k=1,\ldots,n$. Therefore, we can limit ourselves to
considering only such strategies.

To enumerate all functions with success probability $\omega_{cl}$
we need $\log M_c(\omega_{cl})$ bits. One of the possible ways of
enumeration is first to describe the optimal strategies. To do so
one needs to describe all the functions $y_k(x_k)$. There are
$2^m$ functions for every $k$, which gives $2^{mn}$ sets of $n$
different functions. Then one can calculate the outputs of this
strategy $B=(y_1(x_1),...,y_n(x_n))$. To fully characterize $f$
one can simply take the description of the classical strategy,
which requires $nm$ bits, and append to it a list of numbers
$\vec{b}(\vec{x})=f(\vec{x})\ominus B(\vec{x})$. Where $\ominus$
is bitwise subtraction modulo $m$. However, each of numbers
$\vec{b}(\vec{x})$ is equal to 0 whenever the classical strategy
succeeds, which happens with probability $\omega_{cl}$. There are
$m^n$ numbers $\vec{b}$ which need to be specified and each of
them has $2^n$-letter alphabet, but because 0 will appear with
probability $\omega_{cl}$ the entropy of the whole set of
$\vec{b}$'s is at most $h^*(\omega_{cl}) m^n$, where
$h^*(\omega_{cl})=h(\omega_{cl})-(1-\omega_{cl})\log (2^n-1)$. $h$
is Shannon's binary entropy and $h^*$ is the highest entropy a
variable with $2^n$-letter alphabet can have if one of the letters
has probability $\omega_{cl}$. It can be considered a
generalization of $h$. $h^*(\omega_{cl})$ is equal to $n$ only if
$\omega_{cl}=2^{-n}$ and is strictly smaller otherwise.

This encoding uses \be M'=mn+h^*(\omega_{cl})m^n \ee bits. Since
any encoding must require at least $\log M_c(\omega_{cl})$, then
$M'\geq \log M_c(\omega^*_{cl})$, which gives an upper bound on
$M_c(\omega^*_{cl})$. An important implication is that for any
$\omega_{cl}>2^{-n}$ the ratio $\frac{M_c(\omega_{cl})}{M}$ goes
to 0 as $m$ goes to infinity. This means that as the size of the
input grows the fraction of functions where the classical
resources provide a non-negligible advantage over producing random
outputs goes to zero which proves the claim.



We actually proved more: for a majority of the functions
non-signalling resources can achieve a success rate $\approx 2$
times higher than the classical ones.

\section*{DISCUSSION}

In this work, we addressed the following question: {\it is quantum
mechanics superior to classical theory regarding distributed
computation of total function, if no communication between nodes
is allowed?} In order to answer this question we introduced Id
games. The motivation for the studies of these games comes from
the fact that they are instances of a complete nonlocal
computation. In the case of any type of distributed computation
any exclusions of some possible combinations of inputs for the
parties are difficult to justify. Therefore, the functions $g$
that we consider are total functions. Moreover, function $g$ is
defined such that only one output value $\vec y$ is a valid answer
for a given input $\vec x$. This research has also been partially
inspired by the results of \cite{gyni} where the authors study a
particular family of inequalities that is similar to Id games but
defined for partial functions {(i.e. there is a promise on the players' inputs)}. The results from \cite{gyni} have been very useful in the studies of the foundations of the quantum
theory \cite{multi,LO} and we believe that Id games will too.


\section*{Acknowledgements}

This work was supported by  the ERC AdG grant QOLAPS, EC grant
RAQUEL (323970) and National Science Centre project Maestro DEC-2011/02/A/ST2/00305. M.P was supported by NCN grant 2013/08/M/ST2/00626 and FNP TEAM programme.



\end{document}